\documentclass[10pt, conference, compsocconf]{IEEEtran}

\ifCLASSINFOpdf
  \usepackage[pdftex]{graphicx}
\else
  \usepackage[dvips]{graphicx}
\fi

\usepackage[cmex10]{amsmath}
\begin{document}

% paper title
\title{An Ising Model Approach to Malware Epidemiology}

% author names and affiliations

\author{\IEEEauthorblockN{Kristine Eia S. Antonio\IEEEauthorrefmark{1}\IEEEauthorrefmark{3}, Chrysline Margus N. Pi\~nol\IEEEauthorrefmark{1}\IEEEauthorrefmark{2} and Ronald S. Banzon\IEEEauthorrefmark{1}}
\IEEEauthorblockA{\IEEEauthorrefmark{1}Structure and Dynamics Group,
National Institute of Physics \\ UP Diliman, Quezon City, Philippines 1101}
\IEEEauthorblockA{\IEEEauthorrefmark{2}Institute of Mathematical Sciences and Physics \\ UP Los Ba\~nos, Laguna, Philippines 4031}
\IEEEauthorblockA{\IEEEauthorrefmark{3}kantonio@nip.upd.edu.ph}}

% make the title area
\maketitle

\begin{abstract}
We introduce an Ising approach to study the spread of malware.  The Ising spins up and down are used to represent two states--online and offline--of the nodes in the network.  Malware is allowed to propagate amongst online nodes and the rate of propagation was found to increase with data traffic.  For a more efficient network, the spread of infection is much slower; while for a congested network, infection spreads quickly.
\end{abstract}

\begin{IEEEkeywords}
computer networks; computer viruses; epidemiology
\end{IEEEkeywords}

\section{Introduction}
The internet has become a near indispensable tool with both private individuals and organizations becoming increasingly dependent on internet-based software services, downloadable resources like books and movies, online shopping and banking, and even social networking sites. The issue of network security has become significant due to the prevalence of software with malicious or fraudulent intent. Malware is the general term given to a broad range of software including viruses and worms designed to infiltrate a computer system without the owner's permission \cite{Hao}\cite{ISBS}. Cohen's conclusion in his 1987 paper that computer viruses are potentially a severe threat to computer systems \cite{Cohen} is still valid in real networks today \cite{ISBS}\cite{Chakrabarti}\cite{prosecurity}.  Current security systems do little to control the spread of malicious content throughout an entire network \cite{Chakrabarti}\cite{Kephart}. Most security systems are designed to protect a single computer unit. These properly protected units make up only a fraction of online computers.  These highlight the necessity of examining the dynamics of the spread of malware in order to be able to develop proper control strategies.

Studies on the spread of malware in computer networks date back to the late 1980s \cite{mcgill} and are generally based on the mathematical approach to the spread of diseases in biological populations.  Math models developed for spread of malware within a computer network such as the Kephart-White model and other models adapted from it are based on the Kermack-McKendrick model.  These models have an implicit assumption that all nodes in the network are always available for ``contact'' \cite{Chakrabarti}\cite{murray}.   However, it is a basic limitation of malware that it can only be passed on to another computer if there is a path through which information can be passed \cite{Cohen}, so the states of the nodes of the network--whether they are online or offline--have an effect on the dynamics of the spread.

In this work, we model the spread of malware utilizing an Ising system to represent an isolated computer network.  The state of each node is a composite of its connection status and health.  The spin state of a node defines its connection status to be either online or offline.  Connections are established with the premise that autonomous networks configure themselves \cite{TaoJiang}.  The health status describes whether a node has been infected or not, and infection can propagate only among online nodes.

The Ising model was originally intended for simulating the magnetic domains of ferromagnetic materials.  Its versatility has allowed it to be applied to other systems wherein the behavior of individuals are affected by their neighbors \cite{Giordano}\cite{Landau}\cite{TaoJiang}.  It has been applied to networks and network-like systems \cite{Landau} such as neural networks \cite{Giordano}\cite{TaoJiang}, cooperation in social networks, and analysing trust in a peer-to-peer computer network \cite{TaoJiang}. 

\section{The model}

A computer network is modeled by an $N \times N$ Ising spin system. Associated with each node is a spin $s_{i,j}$ corresponding to two possible states: $+1$ for online and $-1$ for offline. The local interaction energy is given by

\begin{equation}
\label{generalIsing}
 E_{i,j} = -s_{i,j}J_{i,j}\sum\nolimits s_{\substack{nearest\\neighbors}}.
\end{equation}

\noindent The interaction parameter, $J_{i,j}$, determines the degree and type of dependence of $s_{i,j}$ on its neighbors.  The nearest neighbors or local neighborhood are defined according to the network topology and are usually Von Neumann or Moore neighborhoods \cite{Chopard}\cite{Wim}.  Summing up all local energies gives the total energy, $E$, of the system. Global energy, $E$, is associated with network efficiency and more efficient networks are characterized by lower energies. 

Note that while interaction energies are explicitly dependent on the nearest neighbors, the state of each node is implicitly dependent on the state of the entire system. A node will change its configuration provided that the new energy of the system is lower than the previous. If the resulting energy is higher, the new configuration is accepted with probability

\begin{equation}
\label{probabilityIsing}
 p = e^{-\Delta E/k_BT}.
\end{equation}

\noindent In the standard Ising procedure, $\Delta E$ is the change in energy, $T$ is temperature, and $k_B$ is the Boltzmann constant. Here, $T$ relates to network traffic. 

To model the spread of infection, each node is assigned a health status separate from its spin. The health status is either infected or susceptible. Every online susceptible has a probability $P_{inf}$ of becoming infected, where

\begin{equation}
\label{probabilityInfect}
P_{inf} = \frac{\textrm{number of online infective nodes}}{\textrm{number of online nodes}}.
\end{equation}

\noindent Offline nodes do not transmit or receive data. Hence, they do not participate in the infection part.

\paragraph{Program Specifics}  The computer network is a $10 \times 10$ lattice. Nearest neighbors are defined to be the four adjacent nodes. The interaction parameters are all set to $J_{i,j} = J = +1$. Eq.\ref{generalIsing} becomes

\begin{equation}
 E_{i,j} = -s_{i,j}(s_{i+1,j}+s_{i-1,j}+s_{i,j+1}+s_{i,j-1}).
\end{equation}

\noindent For the interaction energy calculations, circular boundary conditions are imposed. Parameters are scaled such that $k_B=1$. Initially, all nodes are offline ($s_{i,j}=-1$). Every time step, the entire system is swept in a left-to-right top-to-bottom fashion, evaluating each node for a possible change in state.  The mean energy per node $\langle E_{i,j}\rangle$ of each configuration is stored and averaged at the end of the run.

The spread of infection begins with a single infective. At $t=0$, one node is selected at random and infected. As the infection spreads, the number of susceptibles, $S(t)$, and infectives, $I(t)$, for each time step are stored. Because no means for removal of infection is provided, all nodes eventually become infected. It is at this time that the program is terminated.  

\section{Analysis of results}

The model was tested for $T$-values ranging from $T=1.25$ to $T=11.25$.  The infection curves of five trials were averaged for each $T$. The average infection curve was normalized by dividing it by the total number of nodes to get the fraction of infectives $i(t)$.  Because it can no longer be assumed that nodes are always available for connection, a regular decay equation is used to model the fraction of infectives curve. 

A system with $N\!\times\!N$ nodes has $S(t)$ susceptibles and $I(t)$ infectives at time $t$.  Within the time-frame $dt$, the number of susceptibles being converted to infectives is $dS(t)$.  As time passes, $dS(t)$ decreases as the population of susceptibles is exhausted. Thus, the probability of conversion, given by $\frac{dS(t)}{S(t)}$ decreases with time.  In equation form, this is

\begin{equation}
\label{decayEq}
 -\frac{dS(t)}{S(t)} = \beta dt
\end{equation}

\noindent where $\beta$ is the decay constant.  The solution to Eq.\ref{decayEq} is

\begin{displaymath}
 S(t) = S_0e^{-\beta t}
\end{displaymath}

\noindent where $S_0$, the initial number of susceptibles, is just the total number of units in the system.  Using these, the expression for the number of infectives, $I(t)$ may be written as

\begin{displaymath}
 I(t) = N^2(1 - e^{-\beta t}).
\end{displaymath}

\noindent This may be normalized to 

\begin{equation}
\label{model}
 i(t) = 1 - e^{-\beta t}.
\end{equation}

\noindent Note that the actual rate of spread varies with time, and $\beta$ provides a measure of the average rate of spread.

The fits were made using the unweighted Levenberg-Marquardt algorithm of Gnuplot ver.4.2 \cite{gnuplot} initialized with $\beta=0.1$. For consistency, because some runs terminate very rapidly, we consider only the first 50 time-steps. 

\begin{figure}[!t]
\centering
\includegraphics[width=2.5in]{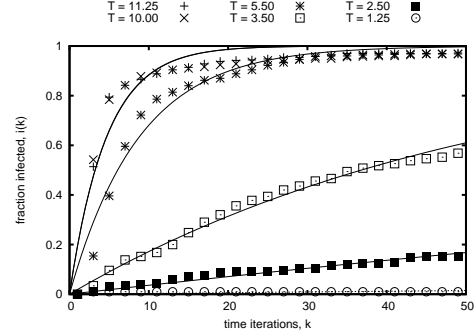}
\caption{\textit{Comparison of Infection Curves for Selected $T$ during the first 50 iterations}: The rate of spread of the infection increases with $T$. For the above graphs, the resulting decay constants are:$\beta(T\!=\!1.25)=0.000150$, $\beta(T\!=\!2.50)=0.003118$, $\beta(T\!=\!3.50)=0.016776$, $\beta(T\!=\!5.50)=0.114230$, $\beta(T\!=\!10.00)=0.214304$, and $\beta(T\!=\!11.25)=0.215791$ }
\label{allcompare}
\end{figure}

From Fig.\ref{allcompare}, it appears that the spread of infection becomes faster as $T$ increases.  For $T = 1.25$ and $T = 2.50$, the rates of spread are very slow, neither reaching $50\%$-infected at the last iteration.  Particularly, for $T = 1.25$, no new infectives were produced.  These low-traffic systems are not dynamic as nodes have a low probability of coming online from their initial offline state.  The network is also very efficient, $\langle E_{i,j}(T\!=\!1.25)\rangle=-4.00$ and $\langle E_{i,j}(T\!=\!2.50)\rangle=-3.57$, which may be interpreted as information exchange being limited to necessary transactions. For this reason, there is little information exchange and hence a slow spread.  For very high $T$, as in $T=10.00$ and $T=11.25$, the spread is rapid and nearly $100\%$ infection is reached.  This suggests that very high traffic means a large volume of information exchange that leads to a faster spread of infection. The system is also inefficient at very high $T$, with $\langle E_{i,j}(T\!=\!11.25)\rangle=-0.76$.  It is worth mentioning that the average infection curves of $T\!=\!10.00$ and $T\!=\!11.25$ nearly coincide indicating rates of spread that are very similar.

\begin{figure}[!t]
\centering
\includegraphics[width=2.5in]{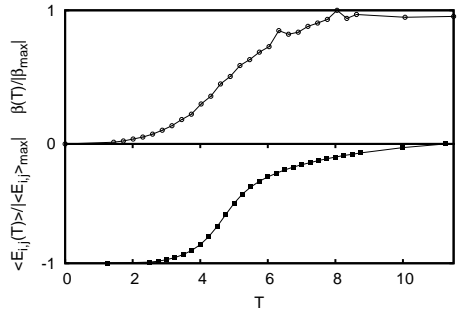}
\caption{\textit{$T$-Dependence of Rates}: The increase in the rate of infection corresponds with the decrease in efficiency in the network.  Note that $E$-values are negative.}
\label{exponents}
\end{figure}

The observations are supported by the calculated decay constants.  The calculated $\beta$ initially increases with traffic but is capped off at very high $T$ where it becomes constant.  This behavior is similar to the saturation region in a traffic network where flux saturates at high densities.  The saturation region indicates that information exchange is no longer freely flowing and that some kind of congestion has occurred \cite{Bing-Hong}.  In Fig. \ref{exponents}, there is an evident transition that occurs in both the average rate of spread and the efficiency of the network.  At the ``congested'' region, the efficiency of the network is very low; while at the ``free flow'' region, the efficiency of the network is comparatively high.  Congestion occurs because networks can only handle a limited amount of traffic--in the form of data packets.  When there is too much traffic, the network is forced to store or drop packets making it inefficient \cite{Bing-Hong}\cite{Fukuda}.  An increase in packet loss with increasing data traffic is reflected by the decrease in efficiency at the congestion region.  The congestion is most likely a result of the limited size of the network and the ``finite-size effect'' may be confirmed by testing a larger network\cite{Bing-Hong}.

\section{Conclusion}
Our Ising model approach accounts for the connection status of nodes in an infected network. Unlike most epidemic models where all nodes are assumed to be always connected, the model allows malware propagation only among online nodes. We found that the rate of infection becomes faster in less efficient networks with higher data traffic and saturates as the network becomes congested.

% use section* for acknowledgement
%\section*{Acknowledgment}

% references section

\end{document}